\magnification=\magstep1
\hoffset=-0.6 true cm
\voffset=0.0 true cm
\baselineskip=12pt

\vsize=8.9truein
\hsize=6.8truein
\tolerance=10000
\parindent=1truecm
\raggedbottom
\def\pp{\parshape 2 0truecm 5.8truein 2truecm 5.01truein}
\def\ltsima{$\; \buildrel < \over \sim \;$}
\def\simlt{\lower.5ex\hbox{\ltsima}}
\def\gtsima{$\; \buildrel > \over \sim \;$}
\def\simgt{\lower.5ex\hbox{\gtsima}}
\def\bline{\hbox to 1 in{\hrulefill}}
\def\etal{{\sl et al.\ }}

\def\vs{{\it vs~}}

\def \wcen{{$\omega~Cen~$}}

\centerline {\bf HST Photometry of 47 Tuc and Analysis of the} 
\centerline {\bf Stellar Luminosity Function in Milky-Way Clusters}
\bigskip
\centerline{Basilio X. Santiago, Rebecca A. W. Elson, Gerard F. Gilmore}
\vskip 0.5 true cm
\centerline {Institute of Astronomy,
Cambridge University, Madingley Road, Cambridge CB3 0HA, United Kingdom}
\bigskip
\centerline{\it Submitted to MNRAS}
\bigskip
\centerline{\bf Key words: Globular clusters: individual (NGC 104) -
Stars: low-mass and statistics}
\bigskip
\bigskip
{\bf ABSTRACT}
\medskip
We present V and I photometry for over 1000 stars in a region 5' from 
the center
of the globular cluster 47 Tuc. The field was imaged with the 
{\it Wide Field and Planetary Camera - 2} 
as part of the Hubble Space Telescope's Medium Deep Survey key project.
The luminosity function (LF) continually rises in the domain $5$ \ltsima
$M_I$ \ltsima $9$ with a slope $\Delta log \Phi (M) / \Delta M \sim 0.15$
and then drops off sharply. We compare our luminosity function
with that derived by De Marchi \& Paresce for a neighbouring HST field.
The two independent LFs are remarkably similar in the
entire range of luminosities probed ($M_I$ \ltsima $10$). 
Comparisons are also made to other HST LFs derived by several 
authors for both globular and open clusters in the Galaxy.
We use the KS test to assess the significance of the differences found.
The luminosity distributions obtained with HST are 
consistent with being derived
from the same population down to $M_I \sim 9.0$. Beyond that, 
statistically significant variations arise. Globular cluster
LFs also differ according to the prominence of a
plateau in the bright end ($5$ \ltsima $M_I$ \ltsima $6.5$). 
The mass functions are
rather uncertain and sensitive to the mass-luminosity relation.
Different approaches to deriving the 47 Tuc mass function from its
LF lead to markedly different results at the low mass end.
For $M$ \gtsima $0.4 M_\odot$, 
the 47 Tuc mass function is significantly
different from that of \wcen.
The calibrated 
HST $M_V(V-I)$ colour-magnitude diagrams (CMDs) show a trend with metallicity
in the expected sense of metal richer systems having redder CMDs for a
fixed absolute magnitude. The main-sequence slope becomes
shallower with increasing metallicity.
The CMDs derived from HST are in
general agreement with previous ground-based studies, 
especially for metal rich stars.
However, the CMDs of metal poor subdwarfs observed from the ground
are shallower than
those of globular clusters observed with HST. This discrepancy may either
result from calibration problems of HST data or reflect real differences
between the CMDs of globular cluster and field halo stars.
\vfill\eject
{\bf 1 INTRODUCTION}
\medskip
The determination of a Globular cluster's luminosity function 
is a first step towards deriving
its present day mass function (PDMF). The latter in turn contains potential
information about the initial
mass function (IMF) and about the role played by 
several dynamical processes such as dynamical friction, evaporation and
tidal stripping over a cluster's lifetime (Elson, Hut \& Inagaki 1987,
Aguilar 1993). 
Knowledge of the low-mass end of the IMF may help modeling star formation
processes and the physics of stellar interiors and atmospheres
(Larson 1992, D'Antona 1994).
Given the similarity between the kinematical and chemical properties
of globular clusters and the stars in the halo of our Galaxy, 
the study of the globular cluster
luminosity and mass functions may also
provide clues to the formation and evolution
history of this latter component.

Derivation of a cluster PDMF from luminosity functions requires knowledge of
the relation between mass and luminosity for its component stars.
Direct mass determinations are available for relatively few disk binaries, 
mostly nearby and of solar abundance (Popper 1980).
Thus, given that the mass-luminosity relation 
is a function of metallicity and age,
mass-luminosity relations must usually rely on models
of stellar atmospheres  
(D'Antona \& Mazzitelli 1986, D'Antona 1987, Bergbusch \&
VandenBerg 1992, Saumon \etal 1994). Such purely theoretical
predictions suffer from substantial uncertainties, especially
for low-mass low-metallicity stars, for which relatively
few high signal-to-noise spectra have been obtained in order to
constrain the theory.

On the observational side, reliable derivation of stellar LFs down to
faint absolute magnitudes ($M_V \sim 14$) has been mostly
restricted to field stars (Wielen \etal 1983, Stobie \etal 1989).
Most previous determinations of the luminosity function (LF) 
in globular clusters
have been based on observations made from the ground, thus suffering
from crowding effects. Fahlman \etal (1989)
and Richer \etal (1991) found evidence of steeply 
rising mass functions at the
low-mass end of several globulars. Their ground-based data, however, required
large and uncertain completeness corrections at the low-luminosity (mass) end.
More recently, data from the Hubble Space Telescope (HST) have become available
as the result of several independent studies (Paresce, De Marchi \& Romaniello
1995, Elson \etal 1995, De Marchi \& Paresce 1995a,b). Given the 0.1'' 
resolution of the {\it Wide Field and Planetary Camera 2} (WFPC-2), 
these HST data allow globular cluster
luminosity functions to be pushed towards much fainter magnitude levels
without the need for large completeness corrections. 
Most of the LFs derived with HST show a clear decline at faint magnitudes,
in conflict with the ground-based work. There also seems to
be a general agreement in the shape of the LF at bright and 
intermediate luminosities, even when
systems of completely different chemical abundances, ages and dynamical
histories are compared (such as the disk field stars \vs 
halo globulars, see Elson \etal 1995). It is thus 
important to confirm this result 
using the increasing amount of available 
HST data, and to quantify the possible differences in the low luminosity end.

In this paper we present data obtained for
47 Tuc as part of the Medium Deep Survey HST key-project 
(MDS, see Griffiths \etal 1994).
We compare the derived colour-magnitude diagram (CMD) and luminosity
function to those obtained by several other authors using HST.
We assess the differences in the observed LFs by means of
the Kolmogorov-Smirnov (KS) test.
In spite of the lack of knowledge about the mass-luminosity relation and its
dependence on metallicity (especially for low mass stars), 
we also attempt to derive the PDMF for the 47 Tuc data and
to compare it to the PDMF of other clusters.
In \S 2, we present the data and discuss the accuracy of our
photometry. In \S 3 we show the LF for 47 Tuc and compare it to those
of other clusters. In \S 4, we do the same analysis for the mass functions.
and also present CMDs. Finally, in \S 5 we present our conclusions.
\bigskip 
{\bf 2 THE DATA}
\medskip\nobreak
{\bf 2.1 The field and sample selection}
\medskip\nobreak
The WFPC-2 field was observed on Sep. 29 1994 in parallel mode.
It is located 4.8' east and 1.3' north of the cluster center,
at a distance corresponding to about $0.8~r_h$, 
where $r_h$ is the half-mass radius (Meylan 1989).
A total of 8 exposures was made with both HST wide I (F814W)
and V (F606W) filters. All frames were put through the standard 
pipeline procedure  which accounts for dark current
and bias level subtraction, flat-fielding and correction for several other
instrumental effects (Holtzman etal 1995a).
The first 4 frames taken with each filter
are spatially offset to the south by
25'' relative to the last 4 frames. 
This offset is large enough to lead to a significant loss of area if
all 8 exposures are coadded together.
Thus, we decided to coadd each set of 
4 exposures separately; the frames in each set were scaled to
a common exposure time and the lower median pixel intensity
used at each position. This method has been successfully applied to HST images
in several previous works (Glazebrook etal 1995, Elson \etal 1995, 
Elson \& Santiago 1995,1996, Santiago \etal 1996) and we are indebted to
K. Glazebrook for developing the IRAF script that performs this task.
The total exposure time for each set was 4000s in the I and 3200s in the V
band. Hot pixels were not 
eliminated during the coadding phase and were not flagged out
either, but were eliminated subsequently (see below).
No correction was applied for the charge transfer
efficiency (CTE) problem, since our data were taken at a CCD temperature of 
$-88^{\circ}$, in which case this is a small effect in the photometry (Holtzman
\etal 1995b).
The two final coadded images in each filter overlap over more than 2/3 of the
3 WFC-2 chips. The WFC chips are also more affected by crowding, since their
spatial resolution is smaller. We thus restricted our analysis to the
smaller field, higher-resolution PC chips, where the overlap region
represents 22\% of the CCD area. We used the objects in this common region
to assess the accuracy of our photometry. We hereafter refer to the
two distinct but overlapping regions imaged with the PC in both filters 
as Fields 1 and 2.

The IRAF package DAOPHOT was used to obtain object lists and
perform the photometry. We adopted a procedure very similar to that
described in Elson \etal (1995). Our primary object list was obtained 
by running the DAOPHOT task DAOFIND on the F814W coadded images of both
Fields 1 and 2. A total of 3530 sources were 
found by the DAOFIND task in Field 1.
Field 2 contributed with an additional 3859 objects. 
Most of the sources detected, however,
turned out to be features around the psf of bright stars or hot pixels.
These features are structurally different from point sources
imaged with the PC chips and were easily eliminated on the basis
of their shape and on their poor fit to the HST-PC point spread function (psf).
The shape of this latter was obtained
by fitting a Moffat function with $\beta = 1.5$ to about 20 bright
and isolated stars in each field.
The resulting psf was then fitted to all sources detected and those with large
$\chi^2$ values or elongated shapes were eliminated. DAOPHOT assigns
a magnitude for each source by scaling the psf template so as to match 
source's intensity profile. Since the psf template's magnitude was obtained 
from aperture photometry on a 2 pixel radius around bright isolated stars,
an aperture correction had to be applied in order to account for the
light outside this radius. This correction was estimated directly
from the same bright isolated stars used to build the psf and showed
very little variation from star to star. It also corresponds to those
quoted by Holtzman \etal (1995a). The aperture corrections
amounted to 0.5 mag in the V filter and 0.6 mag in the I filter.

We visually inspected the remaining objects and almost all of them had shapes
consistent with being stars.
However, some residual non-stellar features 
still remained close to very bright,
saturated stars; they were eliminated by defining exclusion 
zones in the vicinity of such objects. 
Our final stellar lists contained 779
and 763 objects in fields 1 and 2, respectively. 151 of these objects 
($\sim 20\%$) are in common between the two PC fields, very close
to the expected number considering the size of the overlapping area.

For the F606W frames we adopted essentially the same procedure, except for
the fact that we used the same primary object lists as in 
the F814W frames.
The shape of the F606W psf was obtained independently, again by fitting
a Moffat function to several bright isolated stars and creating a template. 
Magnitudes, shape parameters and $\chi^2$ values 
were then obtained for all sources by fitting the template psf.
Using the same criteria as for the I band, 685
sources in field 1 were found to be stars.
Field 2 contributed with 671 stars in V, 135 in common with
Field 1. 
\medskip\nobreak
{\bf 2.2 The photometric accuracy}
\medskip\nobreak
Figure 1 shows a comparison between the
magnitudes obtained for the stars in
common between fields 1 and 2. 
The zero-points applied to these magnitudes were those
listed by Holtzman \etal (1995a) in order to transform the
WFPC-2 flight data into the WFPC-1 system defined by Harris \etal (1991).
We did not apply any colour terms at this stage, since they are 
small ($< 0.1~ mag$). 
The agreement between the two sets
of magnitude measurements is excellent for both filters. 
The scatter in panel {\it a} is
less than $0.04$ for $I_{814} < 20.5$, increasing continuously
towards faint magnitudes up to $0.14$ for $I_{814}$ \gtsima $22.5$.
Similar values are found for the scatter in the $V_{606}$ magnitudes.
There is a very small ($\sim 0.04$) mean offset between the magnitudes
in the two fields, in the sense that field 2 values are fainter. This is
consistent with the amplitude and direction of the CTE effect in the HST
chips, since the common stars between the two fields are located at
high (low) row numbers in field 2 (field 1). Given its small amplitude,
we did not correct for this offset.

In Figure 2 we show the colour-magnitude diagram for the 
composite data of fields 1 and 2. 
Only objects that were successfully classified as stars in both
I and V frames are included. Stars brighter than $I_{814}= 18.5$ were 
eliminated to avoid saturation problems. A total of 1121 stars are shown in the
figure. There are two very distinct
loci in the diagram. Most stars lie along the 47 Tuc main sequence but
many are seen on the lower left of the plot. These later belong to
the Small Magellanic Cloud (SMC) and can be clearly separated from the
cluster stars. The main sequence is fairly well defined, showing a scatter in
colour smaller than that of De Marchi \& Paresce (1995b), whose data correspond
to a nearby field also imaged with HST. This is consistent with our larger
exposure times. 
The broadening of the main-sequence at $I_{814} < 19$ is due to 
residual saturation effects.
Even though most saturated stars were automatically
thrown away because of their poor fits to the PC psf, some objects whose peaks
are slightly saturated have remained. There is also
a hint of a main-sequence curvature
at the bright end, which is caused by the turn-off at $V_{606} - I_{814}
\sim 0.5$.

From Figure 2, we can also identify a parallel sequence of stars lying above
the main sequence. These objects are consistent with being unresolved 
equal mass binaries, whose composite magnitude would thus lie 0.75 mag
above the main sequence. Based on their number we estimate the equal mass
binary fraction at this radius 
in 47 Tuc to be $\sim 5\%$. Finally, 
we also confirm the steepening
in the main-sequence (at $I_{814} \sim 21.5$),
as noticed by De Marchi \& Paresce (1995b) for the same cluster.
This observed ``kink'' is related to changes in stellar interiors
which affect the temperature-luminosity relation (Copeland \etal 1970).
\medskip
\bigskip 
{\bf 3 LUMINOSITY FUNCTIONS}
\medskip\nobreak
The I band luminosity function derived from the 1050
stars found to lie along the main sequence of Figure 2 is shown in Figure 3. 
We assumed a distance modulus of 
$(m-M) = 13.46$ for 47 Tuc when converting $I_{814}$ into absolute magnitudes 
(Madore 1980).
We also applied an extinction correction of $A_{814}=0.08$, consistent
with $E(B-V)=0.04$ (Hesser \etal 1987) and Table~12B of Holtzman \etal (1995b).
The error bars in the figure are Poissonian 
and the luminosity function is expressed
as number of stars per unit magnitude. One can estimate the slope of
the LF brighter than the turn over, where completeness 
corrections are likely to
be irrelevant. We obtain $\Delta \Phi / \Delta M_{814} \simeq 0.15$
for $M_{814} < 9$. This is in excellent agreement with De Marchi
\& Paresce (1995b).

The effect of incompleteness was assessed
with the help of simulations. We used the ADDSTAR task in DAOPHOT
for that purpose. Five simulations with 200 artificial stars 
each were made for each one of 16 
magnitude bins within the range $19 < I_{814} < 27$. The stars were added to
the real data frame and the resulting image was put through the same 
object detection
and star selection procedure described above. 
Magnitudes were also obtained by psf fitting,
as for the real data. For each simulation,
we estimated the fraction of artificial stars recovered and whose magnitudes
agreed within 0.5 mag with the input ones. 
Notice that the completeness function derived this way 
quantifies the loss of stars
solely due to our detection threshold in the F814W frames. 
We are, however, restricting
our analysis to those stars for which a $V_{606}$ magnitude was also available.
In order to incorporate the loss of stars caused by this additional 
constraint, the 
completeness function derived from the simulations had to be multiplied
by the fraction of main sequence stars detected in $I_{814}$ for 
which a $V_{606}$ magnitude was successfully measured, $f_V(I_{814})$. 
This latter quantity is not easy to
quantify, since we do not know what the contamination of SMC stars
at each $I_{814}$ magnitude is in the original object list.
We derived values for $f_V(I_{814})$ using 
two extreme assumptions: {\it a)} completely overlooking the 
presence of SMC stars in the original object list 
and computing $f_V(I_{814})$ as the ratio of
the number of main sequence objects shown in Figure 2 to the total detected
in the $I_{814}$ frames. This 
leads to an underestimate of $f_V(I_{814})$; {\it b)} assuming
that the ratio of main-sequence/SMC stars in the original list
is the same as in Figure 2,
which tends to overestimate $f_V(I_{814})$; the MS/SMC ratio in Figure 2
must be smaller than in the original list since the loss of objects due to
lack of V band detection will be more severe 
for faint (red) main-sequence stars
than for faint (blue) SMC stars. The final value of $f_V(I_{814})$ 
was taken to be the mean of these two. Even for our faintest data
bin ($I_{814} \sim 24,~V_{606} \sim 26$), these two extreme values
of $f_V(I_{814})$ agree within 20\%.

The final completeness function is shown in Figure 4
for both fields 1 and 2. The two completeness curves are similar, with the
data being essentially complete for $I_{814} \sim 20.5$ ($V_{606} \sim 21.5$)
and still nearly 50\% complete for 2.5 magnitudes fainter than that. Beyond
$I_{814} \sim 23.5$ ($V_{606} \sim 25.3$)
incompleteness becomes severe, in agreement with the
increasing paucity of faint stars along the main sequence in Figure 2.
The error bars in Figure 4 are the composition of the standard
deviations in the completeness fractions for the 5 simulations carried out
for each magnitude bin and the uncertainty associated to $f_V(I_{814})$. 
Note that field 2 has a slightly
steeper completeness function, consistent with it being slightly more
crowded than field 1.

In Figure 5 we plot the luminosity function corrected for incompleteness.
Its shape is very similar to the uncorrected one (Figure 3), except for
the slower drop in the number of stars at the faint end.
The decreasing trend in the luminosity function of 
Figure 5 for magnitudes fainter than $I_{814} 
\sim 9$, however, is still quite clear. The error bars take into
account both statistical fluctuations and uncertainties in the completeness
corrections.
Also shown in Figure 5 is the 47 Tuc HST data from De Marchi \& Paresce (1995b)
for another field some 5' from the cluster center. The latter LF was normalized
to our LF in the range $5.5 < M_{814} < 8.5$. 
The two curves are remarkably similar over all the common range in $M_{814}$,
adding confidence to the reality of the observed features.

We now consider how the 47 Tuc LF in our field
compares with those of other globular clusters. 
In Figure 6, we plot the completeness corrected 47 Tuc LF
again, along with those derived by Elson \etal (1995) for $\omega Cen$
and Paresce \etal (1995) for NGC 6397; all these
works are based on HST data. We thus avoid the issue of calibration to
the standard (Johnson-Cousins) photometric system and show all the data
in $M_{814}$. For \wcen, we use the raw data from Elson \etal 
(1995) and adopt a distance modulus of (m-M)=13.92, as quoted by 
Madore (1980). This is slightly different from the value used
in Elson \etal (1995). The 2 fields analyzed by these authors
are located at about 12' and 17' from the cluster's center
($\sim 3.2~r_h$ and $\sim 4.6~r_h$, respectively).
The data for N6397 come directly from
Paresce \etal (1995), who studied a field some $0.6~r_h$ from the
center. Both $\omega Cen$ LFs have been 
normalized to the 47 Tuc one in the range $5.5 < M_{814} < 8.5$.
For Paresce \etal (1995), the normalization was within the range
$6.5 < M_{814} < 8.5$. 

All 4 LFs shown in Figure 6 seem to have similar shape within
the region used for normalization ($6$ \ltsima $M_{814}$ \ltsima $8.5$). 
Outside this range significant differences arise between the \wcen
and 47 Tuc data. The \wcen LFs have a clear 
plateau in the range $5 < M_{814} < 6.5$,
whereas the 47 Tuc data rise with nearly constant slope.
At the faint end, 
the $\omega Cen$ LF seems to flatten out, 
whereas the 47 Tuc is clearly declining. 
The N6397 data are in close agreement with
the 47 Tuc LF over the entire range of magnitudes. The same conclusion applies
to the LF obtained by De Marchi \& Paresce (1995a) for M15, which
is not shown in figure 6 for clarity. In fact, as pointed out by these
authors, the N6397 and M15 LFs are remarkably similar, rising to a maximum
at $M_{814} \simeq 8.6$ and dropping-off sharply for fainter luminosities.

How significant are the differences between the \wcen and 47 Tuc LFs?
In order to answer this question we use the Kolmogorov-Smirnov
test. In Table~1, we list the
results of applying the KS test to the 47 Tuc and $\omega Cen$ data.
Columns 1 and 2 list the LFs being compared and column 3 the range in absolute
magnitude ($M_{814}$) used. Columns 4 and 5 give
the number of points within this range 
(including completeness corrections) and column 6
lists the probability that the two LFs are drawn from the same ensemble.
The first two lines show that the data from the two 47 Tuc PC chips lead to
perfectly consistent LFs. We are thus entitled to put the
47 Tuc data together as we have done.
We also compared the two \wcen fields; the results indicate 
that the two \wcen LFs are consistent with each other over most of
the available range in $M_{814}$. However, the two LFs seem to
be significantly different beyond $M_{814} \sim 9.5$, 
which corresponds to the faintest bin in field 1 shown in
Figure 6 (notice that Elson \etal data extend down to $M_{814} \sim 10.4$).

As for the 47 Tuc \vs \wcen comparison, the 
results are more strongly dependent on the domain in $M_{814}$ used and on the
particular \wcen field being considered. \wcen field 1 is marginally
consistent ($P \sim 10-15\%$) with 47 Tuc for $M_{814}$ \ltsima $9$.
If the faintest bins in Figure 6 are included, however, the two datasets
seem to differ at a much higher confidence level. 
The LFs of \wcen field 2 and 47 Tuc are more similar: for $M_{814} < 9.5$, 
$P \sim 30-70\%$, depending on the particular
range in $M_{814}$ used. At the faint end, 
however, the disagreement again becomes
increasingly larger, confirming the visual impression from the figure.
Finally, the differences in the ``plateau region'' at the bright end are 
statistically significant only for the \wcen 1 LF ($P \sim 5\%$). 
For the comparison between \wcen 2 and 47 Tuc, $P > 50\%$ for
$5 < M_{814} < 7$, but the number of objects is significantly reduced.

We have also compared the 47 Tuc and \wcen LFs to that obtained
by von Hippel \etal (1995) for NGC 2477, again using HST.
The latter is an open cluster
of nearly solar metallicity. The same authors also obtained
V and I data for another open cluster, NGC 2420, but the number of points
available is small, rendering the LF rather uncertain. 
The LFs for both N2477 and N2420 are 
shown in Figure 3 of von Hippel \etal 
(1995). The KS results involving N2477 are shown in Table~2.
The conclusions are more uncertain, given the small number of
stars available in the comparison. The range in magnitudes in common is
also restricted since the N2477 data do not extend 
brighter than $M_I \sim 8$. The N2477 data rise out to
$M_V \sim 12.5$ ($M_I \sim 10$) being thus at variance
with the turnover seen in the 47 Tuc LF ($P$ \ltsima $25\%$).
For \wcen {\it vs} N2477 comparison $P$ is consistently larger ($\sim 30\%$).

In brief, variations in shape are observed in the LFs of Milky-Way clusters
at both extremes of the observed range in luminosities, most 
especially at the faint end. For $M_I$ \ltsima $9$, the LFs discussed here 
are still at least marginally consistent with one another.
The shape of the LF does not seem to correlate with metallicity, since the
N6397, M15 and 47 Tuc all have remarkably similar LFs but very
different metallicities (N6397: $[Fe/H] \simeq -1.91$, M15: $[Fe/H] \simeq
-2.26$, 47 Tuc: $[Fe/H] \simeq -0.6$, Djorgovski 1993). Furthermore,
the 2 \wcen LFs are significantly different from those of the other
3 globulars, despite its intermediary metal abundance ($[Fe/H] \simeq -1.6$).
If the mass-luminosity relation 
depends only and smoothly on metallicity, we can expect
the PDMFs not to correlate with metallicity either.
The 5 LFs discussed here were also obtained
at different half-mass radii, ranging from $0.6~r_h$ to $4.6~r_h$.
In this case, dynamical evolution due to both internal and external
processes such as mass segregation and tidal stripping
may have shaped the mass function of each cluster
in different ways. The coupling of variable dynamical
history plus variations in the mass-luminosity relation (especially as
a function of metal content) clearly render as only tentative 
any conclusion regarding 
the differences and similarities among the present and initial 
mass functions of these clusters from their LFs alone.
In the next section we try to derive the PDMF for 47 Tuc and compare it
with that \wcen, assessing the reliability of the PDMFs.

\medskip
\bigskip 
{\bf 4 MASS FUNCTIONS AND COLOUR-MAGNITUDE DIAGRAMS}
\medskip\nobreak
{\bf 4.1 The 47 Tuc mass function}
\medskip\nobreak
Determining the PDMF requires usage of
a mass-luminosity relation. We adopted here two alternative $M_I - mass$
relations: the one shown in Figure 4 
of De Marchi \& Paresce (1995b) and the $M_V - mass$ relation
corresponding to the
14 Gyr, $[Fe/H] = -0.65$ isochrone given by Bergbusch \& VandenBerg (1992),
with $M_V$ magnitudes converted to $M_I$ by using the $M_I(V-I)$ relation for
the sequence of dwarfs and mild subdwarfs of Monet \etal (1992). 
In fact, De Marchi \& Paresce (1995b) used the
same $M_V - mass$ relation but applied a different
conversion from $M_V$ to $M_I$.
Stars as metal 
poor as $[Fe/H] \sim -1$ are likely to be included in the Monet \etal
(1992) sequence, although most should
have nearly solar metallicity; this range in metallicity nicely brackets
that of 47 Tuc.
The $M_{814}$ magnitudes have been converted to $M_I$ using the
calibration proposed by Holtzman \etal (1995b) (see \S 4.2).
Rather than multiplying the LF shown in Figure 5 by the slope of the
adopted $M_I - mass$ relation, we have directly computed the mass for each
star and then binned the data in mass (in solar units). 

The two mass functions derived for 47 Tuc are shown in Figure 7a. 
The squares use the relation provided by De Marchi \& Paresce (1995b).
Since we have less data, our results are noisier. Yet, our mass
function is in qualitative agreement with that 
shown by these latter authors: an increasing mass function down to
0.3 $M_\odot$ and a flattening for smaller masses. Notice, however, that
a slightly increasing mass function cannot be ruled out even at the
low-mass end. This actually also applies to the De Marchi \& Paresce
data. Adoption of the alternative conversion from $M_V$ to
$M_I$ leads to a clearly increasing mass function for 47 Tuc.
Notice that the uncertainties in the $M_V - mass$ relation of
Bergbusch \& VandenBerg (1992) are not reflected in the two mass functions
shown. Incorporation of such uncertainties
would further reduce the reliability of the mass function for low
mass stars. We thus conclude that the low-mass end of the 47 Tuc mass function 
is subject to substantial uncertainty. Similar conclusions were obtained
by Elson \etal (1995) for the \wcen cluster.

Despite the obvious difficulties in determining the mass of low-mass
stars from their I band magnitudes, a fairly reliable computation of the
PDMF can still be made for higher masses (\gtsima $0.4 M_\odot$). In this
range, the two different mass functions of 47 Tuc are in agreement.
In figure 7b we compare the 47 Tuc PDMF to that of \wcen.
The latter was determined by Elson \etal (1995) 
using the $M_I - mass$ relation given by Brewer \etal (1993).
It is clearly much flatter than that of 47 Tuc for $M$ \gtsima $0.3 M_\odot$.
\medskip\nobreak
{\bf 4.2 CMDs and the calibration of HST data}
\medskip\nobreak
In this section we take advantage of the fact that we have HST CMDs for
four clusters spanning a large range in metallicity, and investigate
how reliable the proposed calibration schemes from the HST filters to
the Johnson-Cousins system are. An appropriate transformation should lead
to a trend in the loci occupied by the main-sequence 
CMDs as a function of metallicity
in the sense that more metal rich clusters should have redder main sequence
colours at a fixed absolute magnitude. 
We should also be able to check for the external
consistency of the HST calibration by comparing the calibrated CMDs with those
obtained from ground-based observations using the standard filters.

Figure 8 shows CMDs for \wcen, 47 Tuc, N2420 and N2477 in a sequence of
increasing metallicity. The open clusters have
been calibrated by von Hippel \etal (1995), using the prescriptions given
by Holtzman \etal (1995a). For the more metal poor stars
in \wcen and 47 Tuc, we used the synthetic transformations listed in
Holtzman \etal (1995b). All 4 panels correspond to reddening-corrected
data. From the figure, one can notice a clear 
trend with metallicity in the expected direction: metal rich systems
have redder main-sequences than metal poor ones.
We have also calibrated and fitted the main-sequence locus of M15,
using the fiducial points listed in Table~1 of De Marchi \& Paresce (1995a)
and the transformations of Holtzman \etal (1995b). The M15 data were
dereddened assuming a $E(B-V) = 0.11$ as quoted by the authors.
The main-sequence of M15 is consistently bluer
than that of \wcen, in accordance with the expected trend
in metallicity ($[Fe/H] = -2.26$ for M15, Djorgovski 1993).
M15, \wcen, 47 Tuc and N2477 all have main-sequences well
described by two straight lines, specially in face of the observational
scatter in their CMDs. 
Only the relatively sparse data for N2420 seem to be consistent
with a single straight line. 
Linear fits to the main-sequences of all 5 clusters are shown in the first
part of Table~3. The 4 cases fitted by
a double line show a steepening in their main-sequences,
the slope in the faint range of $M_V$ being larger than that in
the bright end.
There is also a clear trend of both slopes
to become steeper with metallicity.

The HST CMDs have been compared to those
obtained by several authors using ground-based data in
the standard Johnson-Cousins system (Leggett 1992, Richer \& Fahlman 1992, 
Monet \etal 1992, Brewer \etal 1993). 
Linear fits were also made to these latter (or taken from the
original references) and are shown in the second
part of Table~3.
For metal rich systems, the loci of the calibrated HST data are 
in good agreement with those
derived from previous observations; the fits to
the dwarf main-sequences of Stobie \etal (1989) and
Monet \etal (1992) are well bracketed by
those of N2420 and N2477. This is again indication that the HST data have
been correctly converted to the Johnson-Cousins system.

The N3201 data of Brewer \etal (1993), available for $M_V$ \ltsima $8.5$,
also fit nicely into the trend between main-sequence slope and
metallicity followed
by the globular clusters observed with HST; N3201 has a metallicity
$[Fe/H] = -1.3$, being thus more metal rich than M15 and \wcen but
much more metal poor than 47 Tuc. The intercept is, however, too
small, making the N3201 stars $\sim 0.5 mag$ brighter at a fixed colour
than expected from the HST calibration.
The other ground-based CMDs of metal poor stars also show disagreement
with those of the HST globulars. As shown in Table~3,
the subdwarfs CMDs have slopes $a \sim 3.1-3.5$, being much flatter
than those derived from HST data or N3201.
The CMDs
of Leggett (1992) and Richer \& Fahlman (1992) have very similar slopes and
intercepts. On the other hand, the CMD of the subdwarfs in Monet \etal (1992)
has a larger intercept, being about 1 mag fainter at a fixed V-I colour.
This had previously been noted by Richer \& Fahlman (1992), who attributed
the discrepancy to differences in metallicity, Monet \etal subdwarfs
being extreme cases of metal-poor ones. As for the discrepancy 
between the ground-based CMDs and those from the HST, it could indicate
problems in the HST ``synthetic'' calibration for
metal-poor stars. Alternatively, it may reflect
real differences between the CMDs of globular cluster 
and field halo populations. 

\medskip
\bigskip
{\bf 5 CONCLUSIONS} 
\medskip\nobreak
We have presented HST V and I photometry for over 1000 stars located in a field
about 5' from the center of 47 Tuc. The derived CMD and LF are similar
to that shown by another recent work on the same cluster (De Marchi
\& Paresce 1995b); the LF rises continuously down to $M_I \sim 9$, with
a slope $\Delta \Phi / \Delta M_I \simeq 0.15$ and drops-off at
fainter luminosities. The CMD shows a clear and narrow main-sequence
down to $M_V \sim 13$ with a change in slope occurring at
$M_V \sim 9$. A small equal-mass binary main sequence
is also present, amounting to about $5 \%$ of the stars.

By quantitatively comparing several LFs recently obtained with HST, 
we conclude that a universal shape of the stellar LF cannot
be ruled out for $M_I$ \ltsima $8.5-9$ (roughly $M$ \gtsima 
$0.3-0.4 M_\odot$) with the present data. However, a plateau
is seen in the \wcen LF for $5$ \ltsima $M_{814}$ \ltsima $7$ but not
in the other clusters whose LFs are compared. This feature is likely
to be statistically significant, according to our analysis.
Globular clusters with metallicities both higher (47 Tuc)
and lower (M15 and N6397) than that of \wcen do not exhibit this plateau,
suggesting that variations in metallicity are not responsible
for the existence of this feature. A similar conclusion applies to the
differences seen in the low-luminosity end of the LFs: whereas 47 Tuc,
N6397 and M15 show a clear turn-over, \wcen does not.
At the faint end, the differences observed between 47 Tuc
and \wcen are significant with more than 90\% confidence.
Disturbingly enough, significant differences at the faint end of the LF
were found even within the \wcen data alone.
Therefore, we conclude that
the overall shape of the PDMF in clusters is
largely insensitive to metallicity 
unless any metallicity effect is compensated for
by variations in the mass-luminosity
relation so as to lead to similar
shapes for the LF of different metallicities. 
The LF for the open cluster NGC 2477 (von Hippel \etal 1995)
was also compared to those of the 47 Tuc and \wcen, but
the results are not conclusive due to the small number of stars
available in these comparisons. 
We should point out that it is imperative that
LFs of additional clusters with different metallicities are obtained 
to confirm the conclusions presented here.

We have attempted to derive the PDMF of 47 Tuc using
recent mass-luminosity relations. These are, however,
increasingly uncertain at the faint end. We have shown that
uncertainties in photometric transformations alone are enough to
jeopardize any reliable conclusion concerning the mass function
for $M$ \ltsima $0.3 M_\odot$ stars. At higher masses, our limited dataset
allows us to conclude that the derived 
\wcen and 47 Tuc mass functions are significantly different,
the former being much flatter than the latter.
In spite of the uncertainties in the derivation of the PDMF,
it is tempting
to discuss what the best available estimates of the PDMF can teach us
about the IMF and dynamical history of globular clusters.
The PDMFs of N6397 and M15 have been shown to be very similar and
flattened at the low-mass end (Paresce \etal 1995, De Marchi \& Paresce
1995a). The present work and that of De Marchi \& Paresce (1995b)
suggest a flattened PDMF for 47 Tuc as well but with a different shape.
These differences in shape 
between M15 and N6397 on one side and 47 Tuc on the other have been discussed
by De Marchi \& Paresce (1995b).
These authors suggest that they may reflect
metallicity-dependent variations in the IMF 
of these clusters; dynamical effects
would be relatively unimportant at locations not far from the half-mass 
radius and would have been expected to affect the 3 clusters in significantly
different ways. However, the best estimate of the \wcen PDMF by
Elson \etal (1995) does not seem to corroborate this scenario of a metallicity
dependent, flattened IMF, 
with little or no role played by subsequent evolution.
Alternatively, the steep PDMF of \wcen
suggests a steep IMF and a dynamically unevolved system with at best a hint of
depletion of low-mass stars at large radii, as attested by the differences
at the low-mass end between the two \wcen fields studied.
In this case, if we are to assume a universal IMF, the data on N6397, M15
and 47 Tuc would point towards
substantial dynamical evolution in these clusters
even at $\sim r_h$, leading to depletion of
low-mass stars.

Given the spread in metallicity of the available HST data from this
work, Elson \etal (1995), von Hippel \etal (1995) and the works
of De Marchi, Paresce and collaborators, we were able to
investigate any trend in the locus of the main sequence as a function
of chemical abundance. A trend is clearly seen 
in the sense that at a fixed luminosity,
metal rich stars are redder than metal poor ones for a fixed luminosity.
Metal poor CMDs are also steeper than metal rich ones.
Once the data are transformed into the standard photometric system, 
the HST main-sequences seem to agree well with earlier ground-based works,
at least for metal rich systems.
The disk main sequences of Stobie \etal (1989) and 
Monet \etal (1992) are in close agreement with those
of N2420 and N2477 derived using HST. The slope of the main-sequence of N3201,
a globular cluster studied by Brewer \etal (1993), is also in agreement with
the HST calibration, although the intercept is about 0.5 mag too
bright.
We also find a significant discrepancy between the main-sequence fits
to the globular clusters observed with HST and those of ground-based
subdwarfs; these latter have much flatter CMDs than
the former.
These discrepancies may either indicate problems in the
HST calibration for metal poor stars or reflect real
differences between the globular cluster and field halo
stellar populations. 

\medskip\nobreak
\bigskip
{\bf ACKNOWLEDGMENTS} 
\medskip\nobreak
This work was carried out as part of the Medium Deep Survey key HST project.
We thank Ivan King, Ted von Hippel, Pavel Kroupa and Gordon Drukier 
for useful discussions. 
\vfill\eject

{\bf REFERENCES}
\def\pp{\parshape 2 0truecm 13.4truecm 2.0truecm 11.4truecm}
\def\apjref #1;#2;#3;#4 {\par\pp#1, #2, #3, #4 \par}

\medskip
\hyphenation{MNRAS}

\pp Aguilar, L.A., 1993, in {\it Galaxy Evolution: The Milky-Way Perspective},
ASP Conf. Ser., vol. 49, Ed: Steven R. Majewski, p. 155.

\pp Bergbusch, P.A., \& VandenBerg, D.A., 1992, ApJS, 81, 163.

\pp Brewer, J., Fahlman, G., Richer, H., Searle, L., \& Thompson, I., 1993,
AJ, 105, 2158.

\pp Copeland, H., Jensen, J.O., \& Jorgensen, H.E., 1970, A\&A, 5, 12.

\pp D'Antona, F., \& Mazzitelli, I., 1986, A\&A, 162, 80.

\pp D'Antona, F., 1987, ApJ, 320, 653.

\pp D'Antona, F., 1994, {\it in The Bottom of the Main Sequence--And
Beyond}, ESO Workshop, Ed. C. Tinney (Berlin: Springer), in press.

\pp De Marchi, G., \& Paresce, F., 1995a, A\&A, preprint.

\pp De Marchi, G., \& Paresce, F., 1995b, A\&A, preprint.

\pp Djorgovski, S. G., 1993, in {\it Structure and Dynamics of Globular
Clusters}, ASP Conf. Ser. 50, Eds: S.G. Djorgovski \& G. Meylan (San Francisco:
ASP), 3.


\pp Elson, R.A.W., Hut, P., \& Inagaki, S., 1987, ARAA, 25, 565.

\pp Elson, R.A.W., Gilmore, G., Santiago, B.X., \& Casertano, S., 1995, AJ,
110,682.

\pp Elson, R.A.W., \& Santiago, B.X., 1995, MNRAS, 110,682.

\pp Elson, R.A.W., \& Santiago, B.X., 1996, MNRAS, in press.

\pp Fahlman, G. G., Richer, H.B., Searle, L., \& Thompson, I.B., 1989, ApJ,
343, L49.

\pp Glazebrook, K., Ellis, R.S., Santiago, B.X., \& Griffiths, R.E., 1995,
MNRAS, 275, L19.

\pp Griffiths, R., \etal, 1994, ApJ, 435, L19.

\pp Harris, H.C., Baum, W.A., Hunter, D.A., \& Kreidl, T.J., 1991, AJ, 101, 
677.

\pp Hesser, J., Harris, W., VandenBerg, D., Allwright, J., Shott, P., \&
Stetson, P., 1987, PASP, 99, 739.

\pp Holtzman, J.A., \etal, 1995a, PASP, 107, 156.

\pp Holtzman, J.A., \etal, 1995b, PASP, 107, 1065.

\pp Kroupa, P., Tout, C.A., Gilmore, G., 1993, MNRAS, 262, 545.

\pp Larson, R. B., 1992, MNRAS, 256, 641.

\pp Legget, S.K., 1992, ApJS, 82, 351.

\pp Madore, B.F., 1980, in {\it Globular Clusters}, eds: D. Hanes \&
B. Madore, Cambridge: Cambridge University Press, p21.

\pp Meylan, G., 1989, A\&A, 214, 106.

\pp Monet, D.G., Dahn, C.C., Vrba, F.J., Harris, H.C., Pier, J.R., \etal
1992, AJ, 103, 638.

\pp Paresce, F., De Marchi, G., \& Romaniello, M., 1995, ApJ, 440, 216.

\pp Popper, D., 1980, ARA\&A, 18, 115

\pp Richer, H.R., \& Fahlman, G.G., Buonanno, R., Fusi Pecci, F., Searle, L.,
\& Thompson, I., 1991, ApJ, 381, 147.

\pp Richer, H.R., \& Fahlman, G.G., 1992, Nat, 358, 383.

\pp Santiago, B.X., Gilmore, G., Elson, R.A.W., 1996, MNRAS, in press.

\pp Saumon, D., Bergeron, P., Lunine, J.I., Hubbard, W.B., \& Burrows, A.,
1994, ApJ, 424, 333.

\pp Stobie, R.S., Ishida, K., Peacock, J.A.,, 1989, MNRAS, 238, 709.

\pp von Hippel, T., Gilmore, G., Jones, D.H.P., Tanvir, N. \& Robinson, D., 
1995, MNRAS, 273, L39.

\pp Wielen, R., Jahreiss, H., \&, Kr\"oger, R., 1983, in {\it Nearby Stars
and the Stellar Luminosity Function, IAU Colloq 76}, eds: A.G. Davis
Philip, A.R., Upgren, p163. Schenectady, NY: L. Davis.

\vfill\eject

\centerline{{\bf Table~1}. 
$47~Tuc - \omega Cen$ comparison}
\vskip 0.5 true cm
\hrule
\vskip 0.2 true cm
\tabskip=1em plus2em minus.5em
\halign to\hsize
{#\hfil&&\hfil#&\hfil#&\hfil#&\hfil#&\hfil#\cr
LF 1 & LF 2 & Mag. Range & N1 & N2 & P \cr
\noalign{\smallskip\hrule\smallskip}
47 Tuc F1 & 47 Tuc F2 & 5.5-9.5 & 530 & 392 & 0.4403 \cr
47 Tuc F1 & 47 Tuc F2 & 5.0-10.0 & 628 & 485 & 0.7404 \cr
\wcen F1 & \wcen F2 & 5.0-9.0 & 2616 & 713 & 0.3892 \cr
\wcen F1 & \wcen F2 & 4.3-9.5 & 3338 & 911 & 0.5231 \cr
\wcen F1 & \wcen F2 & 4.3-10.0 & 3938 & 1037 & 0.0373 \cr
\wcen F1 & \wcen F2 & 4.3-10.3 & 4572 & 1145 & 0.00005 \cr
47 Tuc & \wcen F1 & 5.0-8.5 & 673 & 2061 & 0.1447 \cr
47 Tuc & \wcen F2 & 5.0-8.5 & 675 & 560 & 0.7168 \cr
47 Tuc & \wcen F1 & 5.0-9.5 & 1059 & 3185 & 0.0390 \cr
47 Tuc & \wcen F2 & 5.0-9.5 & 1058 & 864 & 0.4163 \cr
47 Tuc & \wcen F1 & 5.0-10.0 & 1191 & 3773 &  0.00002 \cr
47 Tuc & \wcen F2 & 5.0-10.0 & 1192 & 993 & 0.1568 \cr
47 Tuc & \wcen F1 & 5.0-7.0 & 265 & 870 &  0.0463 \cr
47 Tuc & \wcen F2 & 5.0-7.0 & 266 & 226 & 0.6599 \cr
}
\vskip 0.2 true cm
\hrule

\vfill\eject

 \centerline{{\bf Table~2}. 
$47~Tuc/ \omega Cen - N2477$ comparison.}
\smallskip
\vskip 0.2 true cm
\hrule
\vskip 0.2 true cm
\tabskip=1em plus2em minus.5em
\halign to\hsize
{#\hfil&&\hfil#&\hfil#&\hfil#&\hfil#&\hfil#\cr
LF 1 & LF 2 & Mag. Range & N1 & N2 & P \cr
\noalign{\smallskip\hrule\smallskip}
47 Tuc & N2477 & 7.9-10.0 & 715 & 26 & 0.2278 \cr
\wcen F1 & N2477 & 7.9-10.0 & 2317 & 26 & 0.3075 \cr
\wcen F2 & N2477 & 7.9-10.0 & 623 & 26 & 0.5516 \cr
}
\vskip 0.2 true cm
\hrule

\vfill\eject

\centerline{{\bf Table~3}. 
Fits to the main sequence of several clusters and field populations.}
\smallskip
\vskip 0.2 true cm
\hrule
\vskip 0.2 true cm
\tabskip=1em plus2em minus.5em
\halign to\hsize
{#\hfil&&\hfil#&\hfil#&\hfil#&\hfil#\cr
Cluster & $[Fe/H]^{1}$ & Range & $a^{2}$ & $b^{2}$ \cr
\noalign{\smallskip\hrule\smallskip}
\noalign{\bigskip {\centerline {HST CMDs}}\medskip}
M15 & $-2.26$ & $6 < M_V < 8.5$ & 5.1 & 2.3 \cr
M15 & $-2.26$ & $8.5 < M_V < 11.5$ & 7.5 & -0.5 \cr
\wcen & $-1.6$ & $6 < M_V < 9$ & 4.6 & 2.5 \cr
\wcen & $-1.6$ & $9 < M_V < 12$ & 5.1 & 1.7 \cr
47 Tuc & $-0.6$ & $6.5 < M_V < 9$ & 3.1 & 3.6 \cr
47 Tuc & $-0.6$ & $9 < M_V < 12.5$ & 4.4 & 1.3 \cr
N2420 & $-0.45$ & $9 < M_V < 14$ & 3.6 & 2.7 \cr
N2477 & $0.00$ & $9 < M_V < 13$ & 3.2 & 2.8 \cr
N2477 & $0.00$ & $13 < M_V < 15$ & 3.6 & 1.6 \cr
\noalign{\bigskip {\centerline {Ground-Based CMDs}} \medskip}
Disk$^{3}$ & $--$ & $10 < M_V < 14$ & 3.5 & 2.8 \cr
Disk$^{4}$ & $--$ & $8 < M_V < 18$ & 3.3 & 2.9 \cr
N3201$^{5}$ & $-1.3$ & $5.5 < M_V < 8$ & 4.3 & 2.1 \cr
Halo$^{3}$ & $--$ & $11.5 < M_V < 14.5$ & 3.5 & 5.4 \cr
Halo$^{6}$ & $--$ & $6 < M_V < 13$ & 3.4 & 4.3 \cr
Halo$^{7}$ & $--$ & $9.5 < M_V < 14$ & 3.1 & 4.5 \cr
}
\vskip 0.2 true cm
\hrule
\vskip 0.6 true cm

\noindent $^1~~Djorgovski~(1993),~von~Hippel~\etal~(1995)$

\noindent $^2~~M_V~=~a~(V-I)~+~b$

\noindent $^3~~Monet~\etal~(1992)$

\noindent $^4~~Stobie~\etal~(1989)$

\noindent $^5~~Brewer~\etal~(1993)$

\noindent $^6~~Richer~\&~Fahlman~(1992)$

\noindent $^7~~Leggett~(1992)$

\vfill\eject

\centerline {\bf FIGURE CAPTIONS}

\item {1-} Comparison between the magnitude 
measurements obtained for the stars in common between 47 Tuc 
fields 1 and 2 
as described in the text. {\it a)} HST I band magnitudes (F814W filter).
{\it b)} HST V band magnitudes (F606W filter).

\item {2-} Colour-magnitude diagram for the HST 47 Tuc field
studied in this paper.

\item {3-} The observed luminosity function of 47 Tuc
obtained from the combination of the PC data in fields 1 and 2. The data are
in the HST F814W system and are not corrected for completeness. Error bars
are Poissonian.

\item {4-} Star completeness functions for fields 1 and 2,
obtained from simulations. The error bars incorporate the standard
deviation of 5 independent realizations at each magnitude bin and the
uncertainty in the quantity $f_V (I_{814})$ as defined in the text.

\item {5-} Solid points: completeness-corrected
luminosity function for
47 Tuc. The error bars incorporate both Poissonian fluctuations in the
number counts and the uncertainty in the completeness function. Open symbols:
the 47 Tuc LF derived by De Marchi \& Paresce.

\item {6-}  A comparison between the 47 Tuc LF
shown in Figure 5 and those derived by Elson \etal (1995) for \wcen and
Paresce etal (1995) for N6397.

\item {7-} {\it a)} The mass function for 47 Tuc
as obtained in two different ways: squares utilize the 
$M_I - mass$ relation given by De Marchi \& Paresce (1995b); triangles
make use of the $M_V - mass$ relation of Bergbusch \& VandenBerg
(1992) and the $M_V$ to $M_I$ conversion of Monet \etal (1992).
{\it b)} The 47 Tuc and \wcen mass functions as derived in the present work
and in Elson \etal (1995), respectively.

\item {8-} Colour-magnitude diagrams for
\wcen (panel {\it a}), 47 Tuc  (panel {\it b}), N2420  (panel {\it c}) and
N2477 (panel {\it d}). The data from the later two panels were taken from
von Hippel \etal (1995) and that in panel {\it a} are from 
Elson \etal (1995). The data were calibrated to the Johnson-Cousins system
using the relations provided by Holtzman \etal (1995a,b). Fits to the main
sequence of each clusters are shown.
\vfill\eject
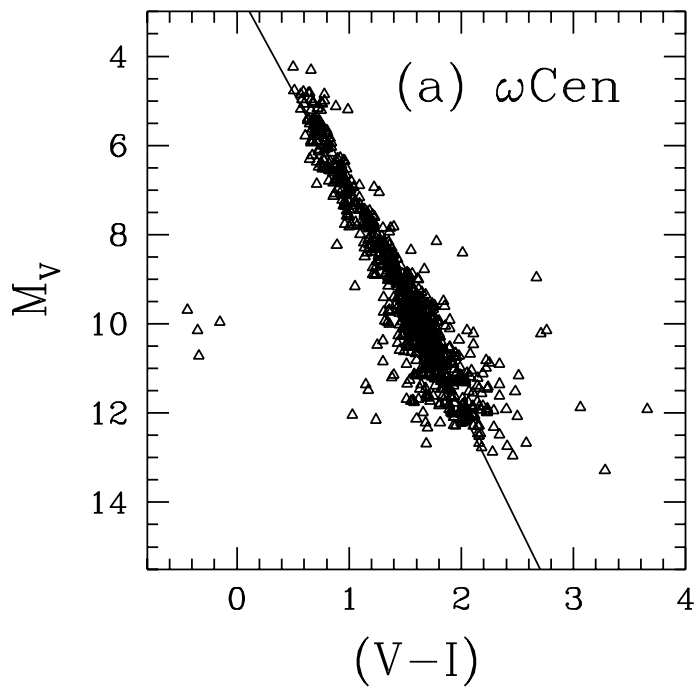
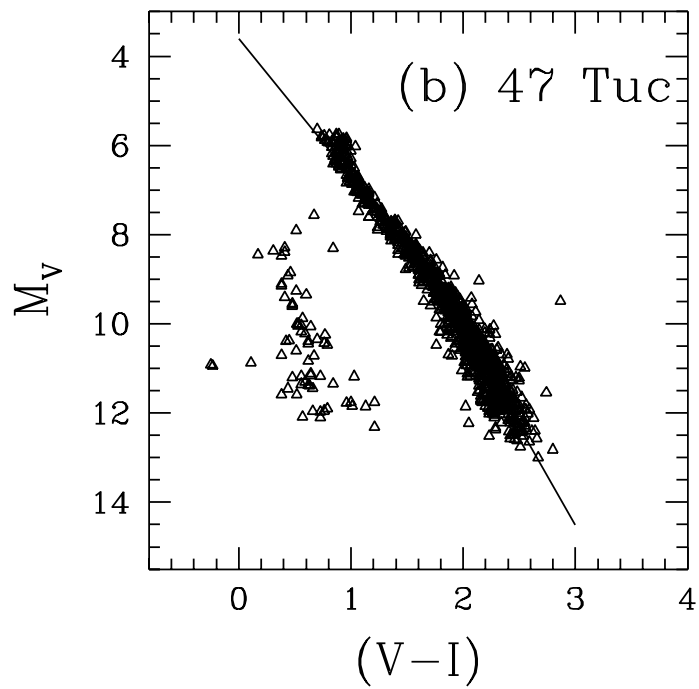
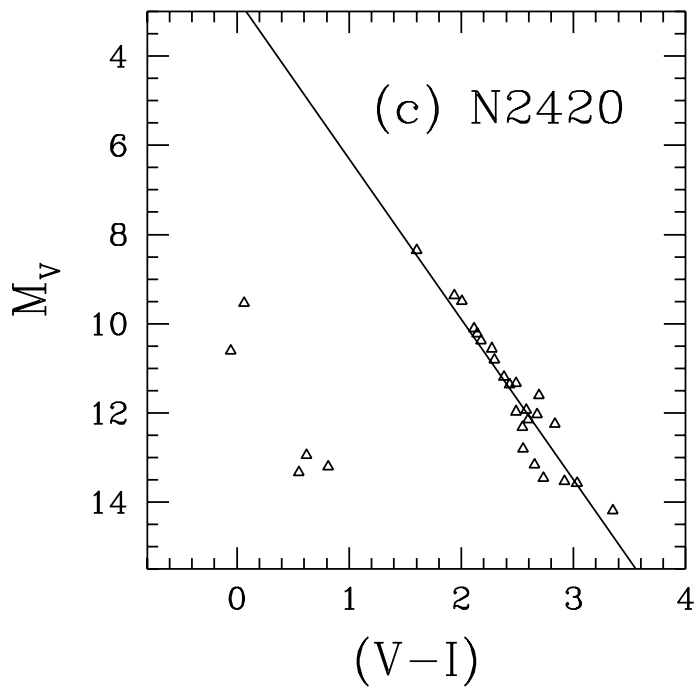
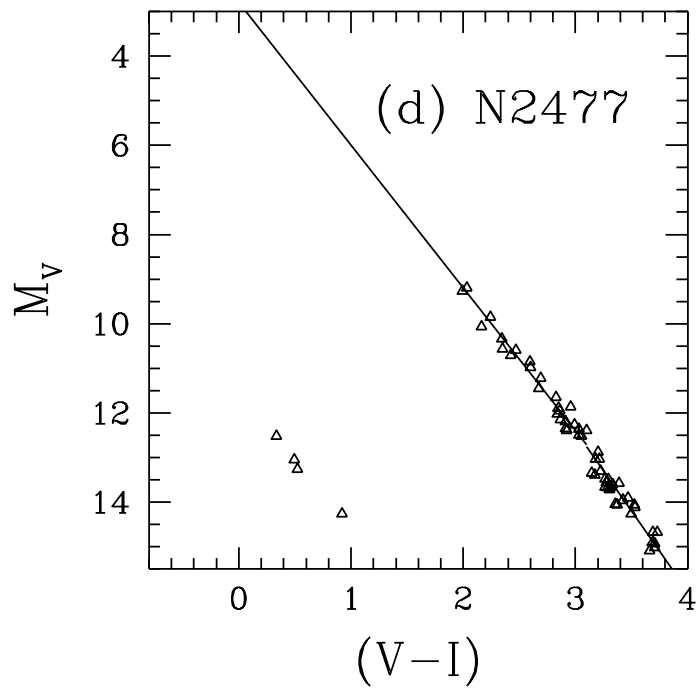
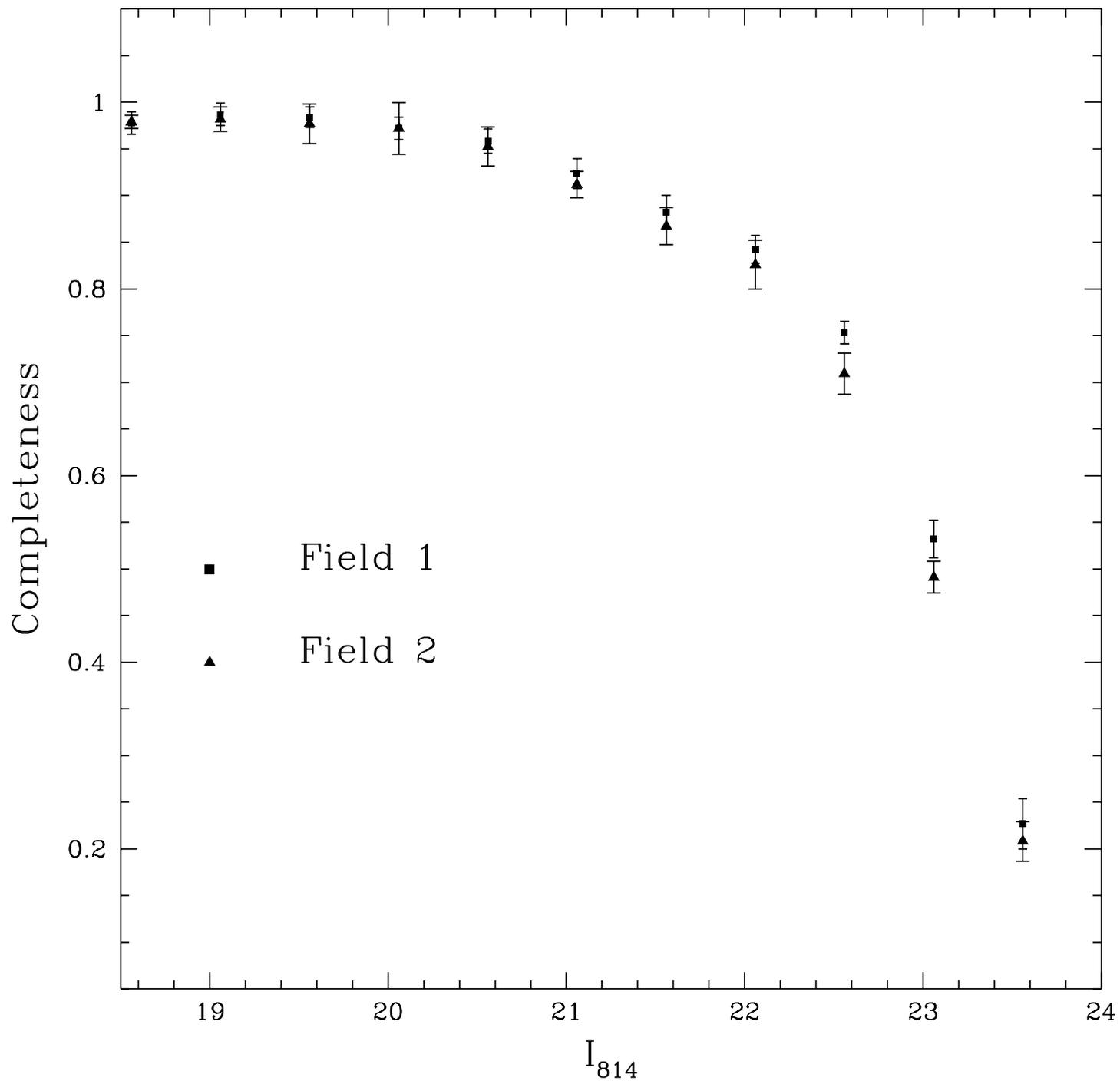
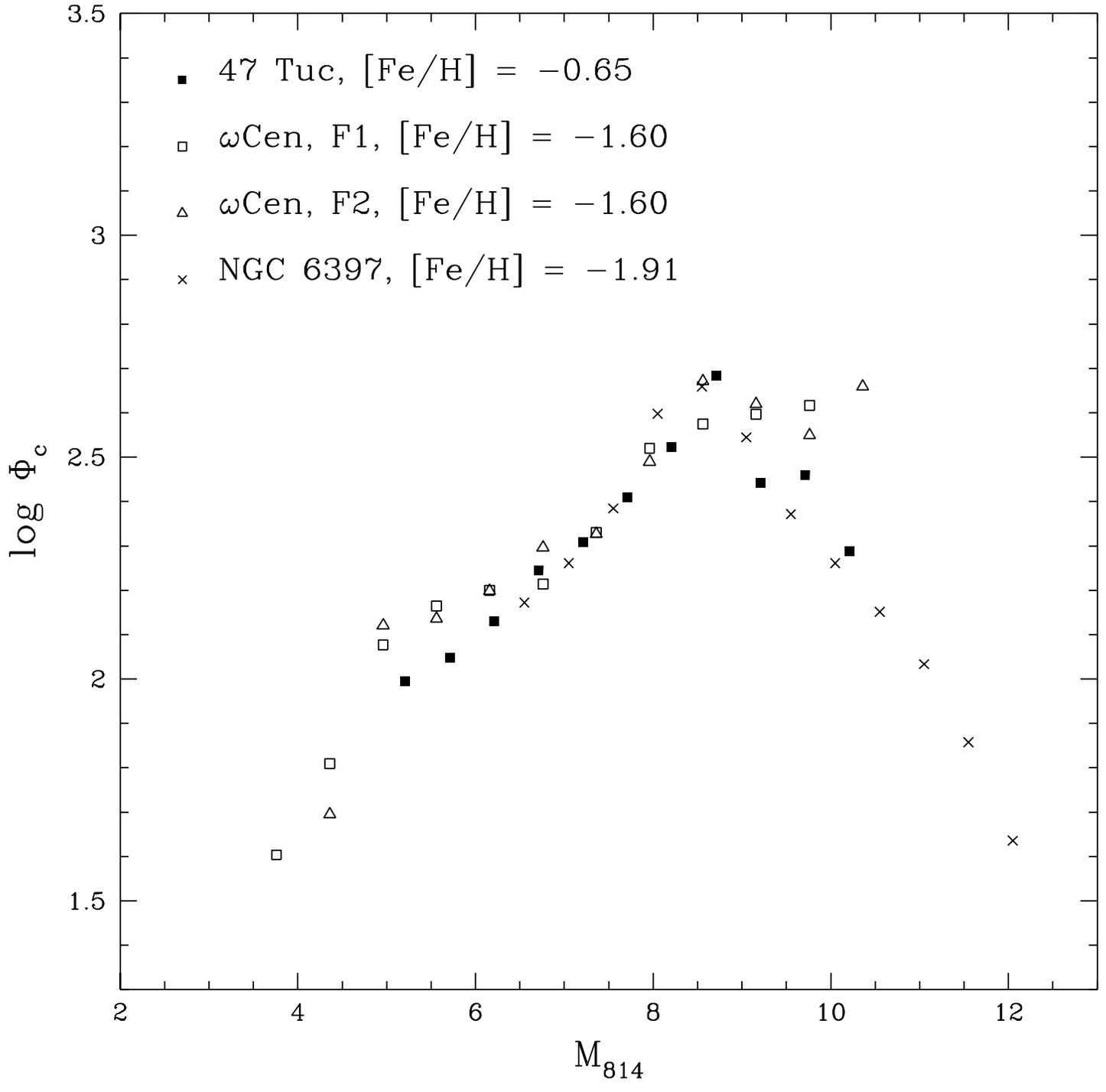
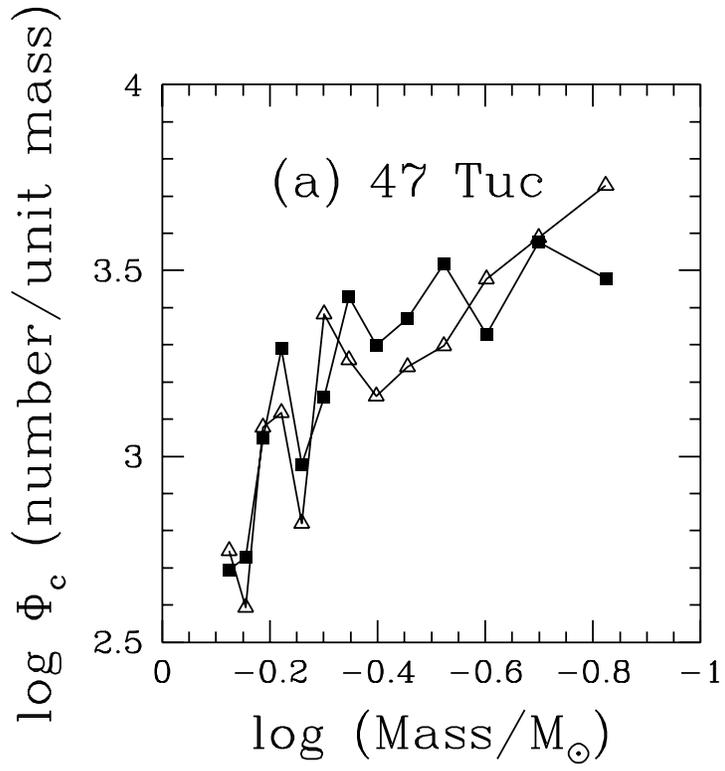
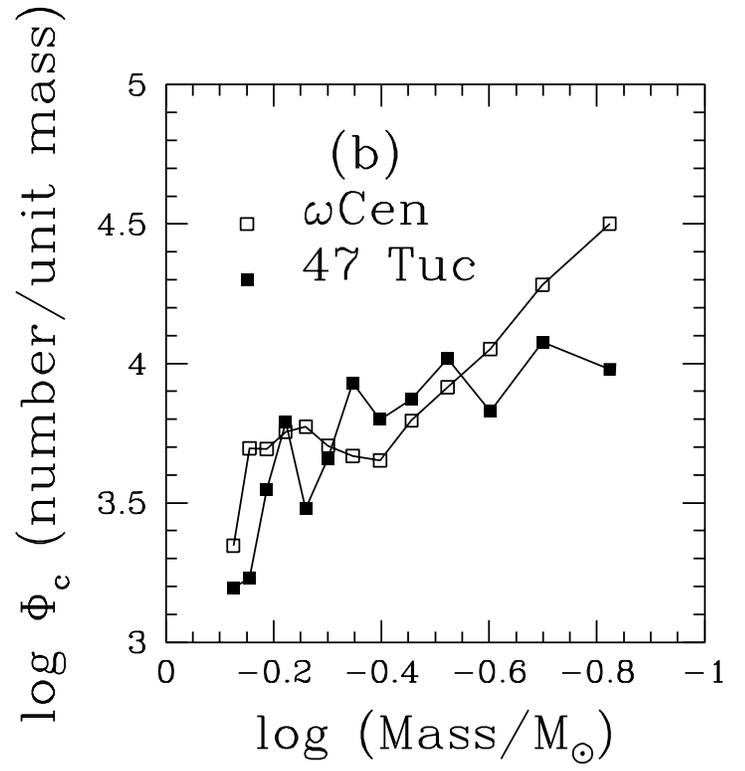
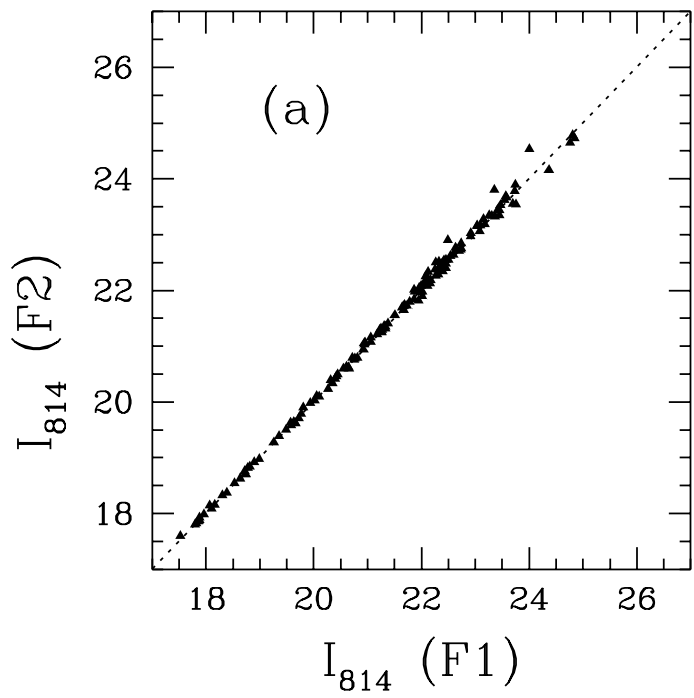
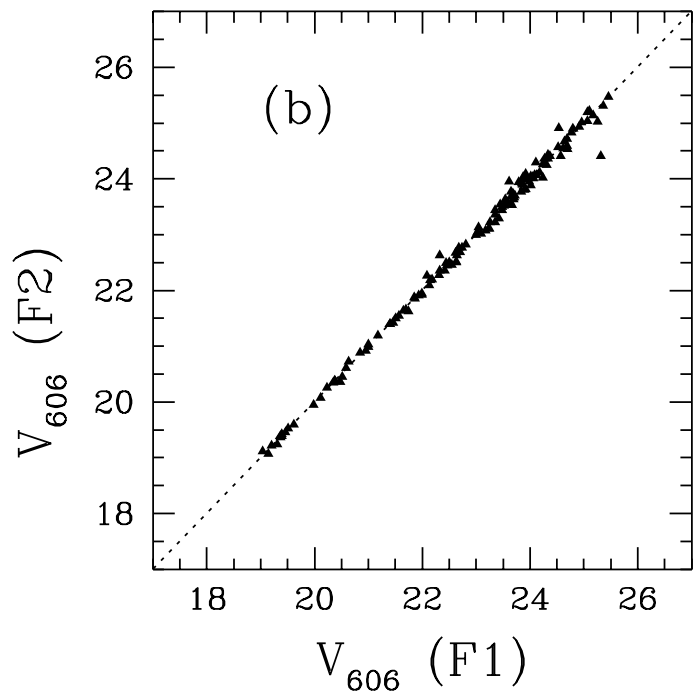
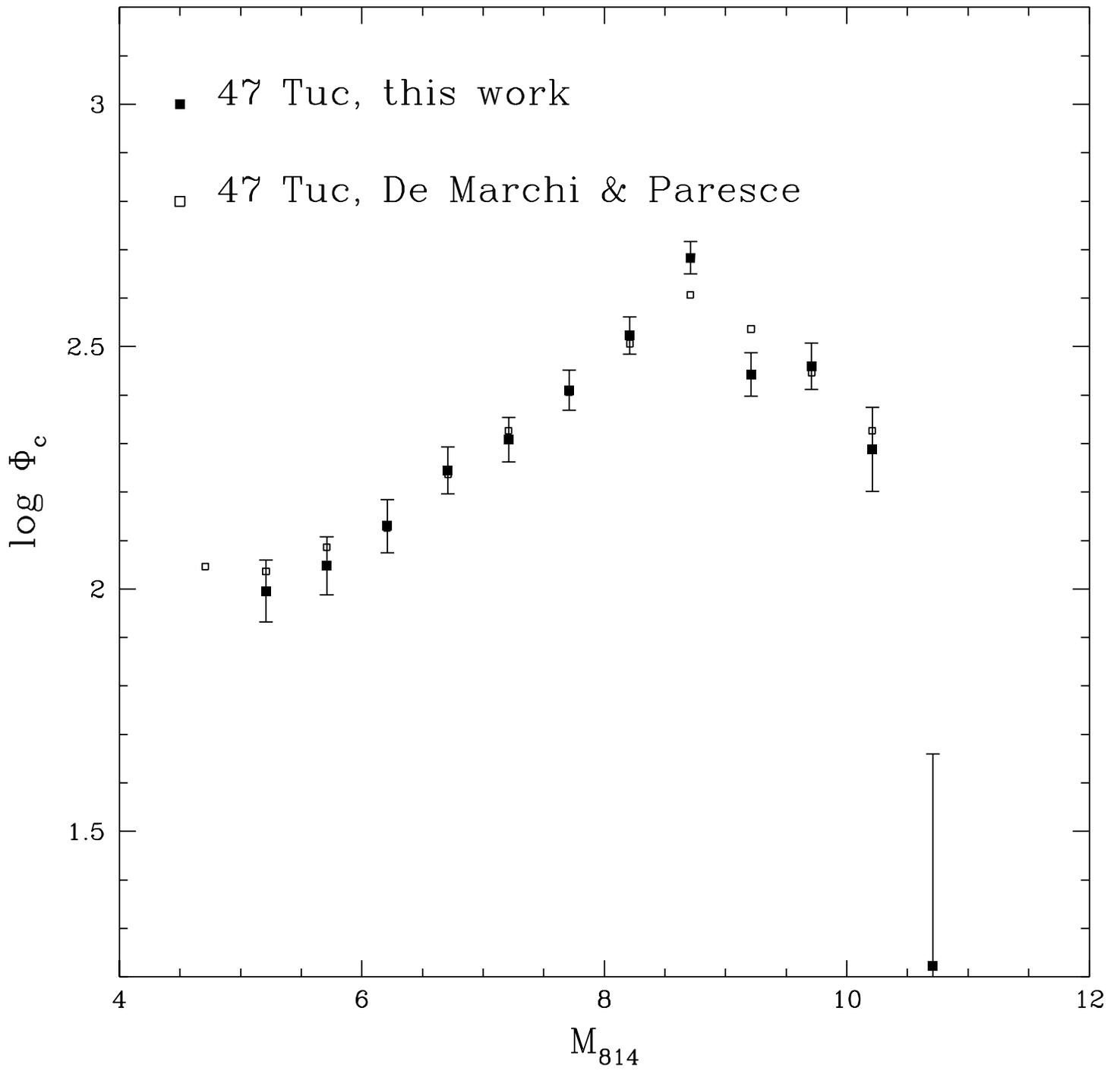
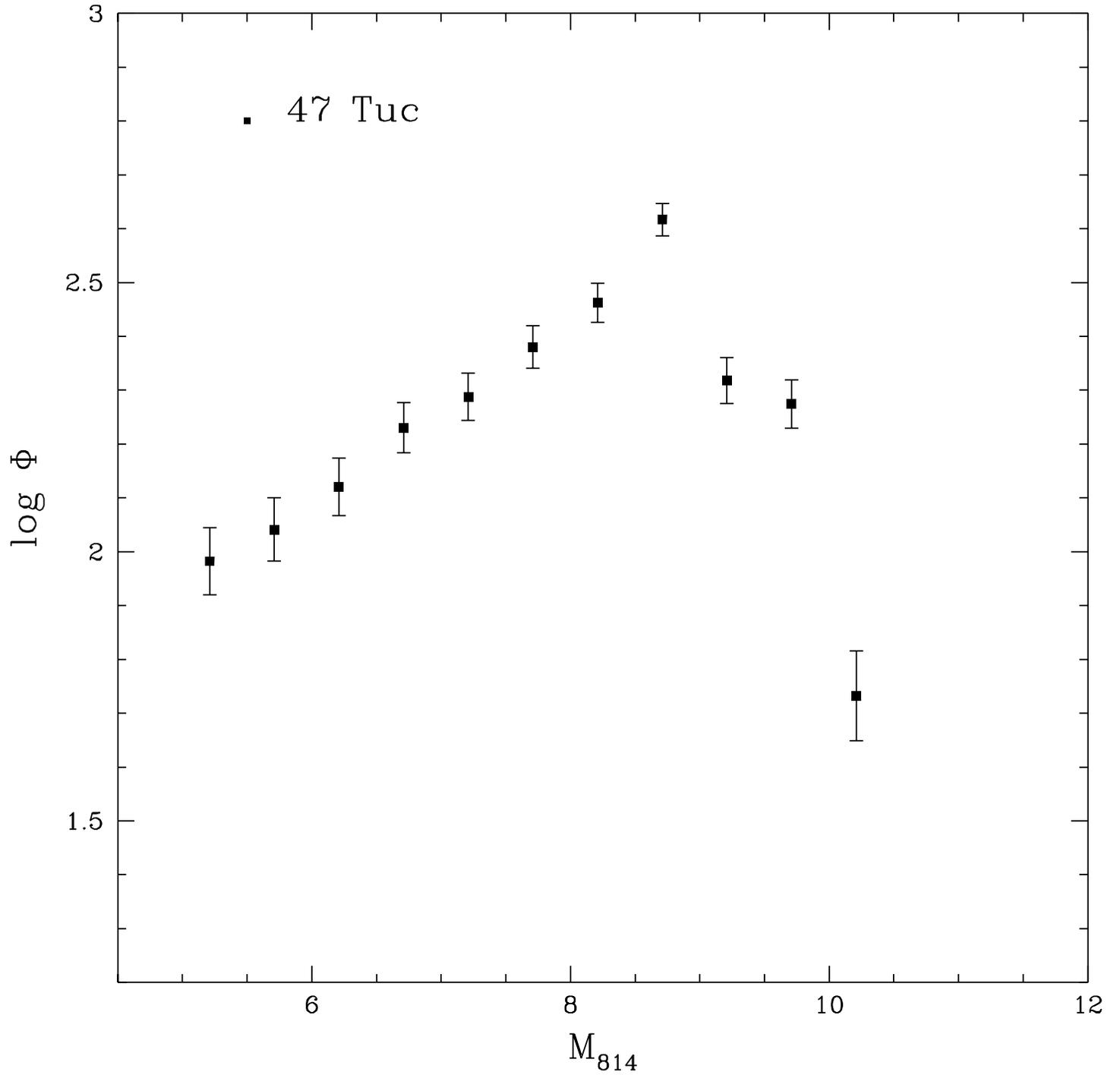
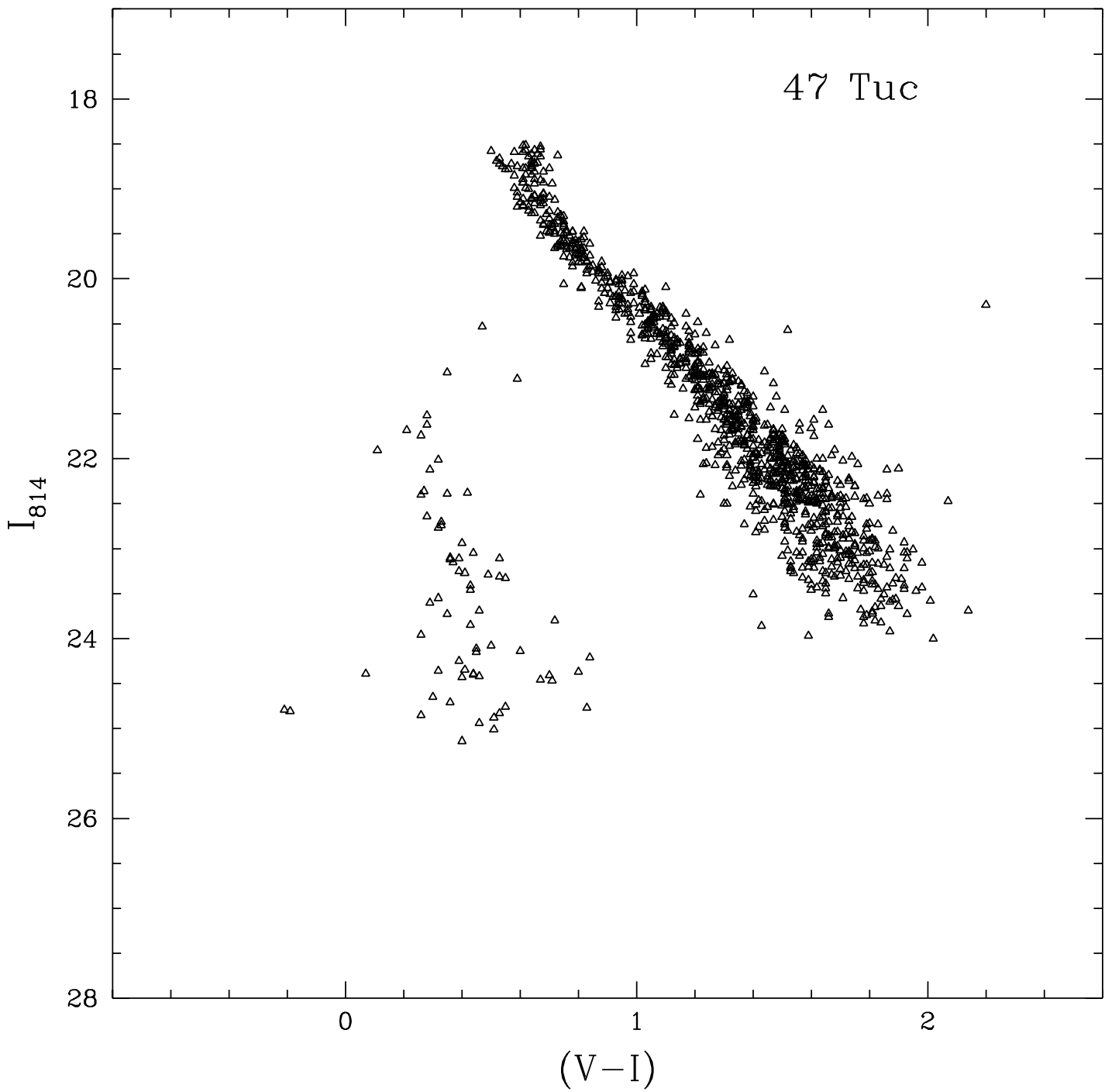
\end